*Article*

# Modeling of photonic integrated resonators using advanced scattering matrix methods

**David J. Moss**

Optical Sciences Centre, Swinburne University of Technology, Hawthorn, Victoria 3122, Australia;
Email: dmoss@swin.edu.au

**Abstract:** We propose a universal approach for modeling complex integrated photonic resonators based on the scattering matrix method. By dividing devices into basic elements including directional couplers and connecting waveguides, our approach can be used to model integrated photonic resonators with both unidirectional and bidirectional light propagation, with the simulated spectral response showing good agreement with experimental results. A simplified form of our approach, which divides devices into several independent submodules such as microring resonators and Sagnac interferometers, is also introduced to streamline the calculation of spectral transfer functions. Finally, we discuss the deviations introduced by approximations in our modeling, along with strategies for improving modeling accuracy. Our approach is universal across different integrated platforms, providing a useful tool for designing and optimizing integrated photonic devices with complex configurations.

## 1. Introduction

With compact footprint and versatile configuration, integrated photonic resonators have become critical building blocks for photonic integrated circuits (PICs), with applications in a wide range of fields such as optical communications [1-3], photonic computing [4-6], nonlinear optics [7-9], sensing [10-12], and optical neural networks [13-15].

Modeling the spectral response of integrated photonic resonators is of fundamental importance for their use in different applications. The scattering matrix method (SMM) (also known as the transfer matrix method) [16, 17], which is derived based on the Maxwell's equations for electromagnetic waves, has been widely employed to model the spectral response of integrated photonic resonators with different device configurations [17-22].

Although in principle the SMM can be applied to model integrated photonic devices with arbitrary planar configurations [16, 23], previous studies mainly focused on modeling devices with simple configurations. A key limitation comes from the fact that the traditional SMM relies on the manual derivation of scattering matrices,, and the obtained scattering matrices need to be multiplied in a specific sequence [24]. For devices with simple configurations, this allows for relatively straightforward calculation of spectral transfer functions. However, for complex integrated photonic resonators, particularly those with bidirectional light propagation, this process becomes much more complicated, which greatly limits its broader applicability.

In this paper, a universal approach based on the SMM is proposed to model integrated photonic resonators with complex structures. The modeling is achieved by dividing a device configuration into basic elements including directional couplers and connecting waveguides, followed by solving a system of linear scattering matrix equations using computational tools. Our approach can be applied to model devices with both unidirectional and bidirectional light propagation, and the simulated spectral response aligns well with measured results for practical devices. The modeling in our approach can also be simplified by dividing the device configuration into independent submodules such as microring resonators (MRRs) and Sagnac interferometers (SIs). Finally, we discuss the limitations of our approach induced by approximations in the modeling, together with strategies for improving modeling accuracy. Our approach provides an effective way for designing and optimizing complex integrated photonic devices across various integrated platforms.

## 2. Modeling of devices with unidirectional light propagation

In this section, we use the device configuration shown in **Figure 1(a)** as an example to show how to model a complex integrated photonic resonator based on the SMM in four steps. The device consists of a microring resonator (MRR) nested within another, with light propagating in only one direction in each waveguide.

First, we divide the device into several directional couplers and connecting waveguides, which are the basic elements in photonic integrated circuits that form integrated photonic resonators. The optical fields at the dividing points between these elements are denoted as $Ei$ ($i$ = 1–16).

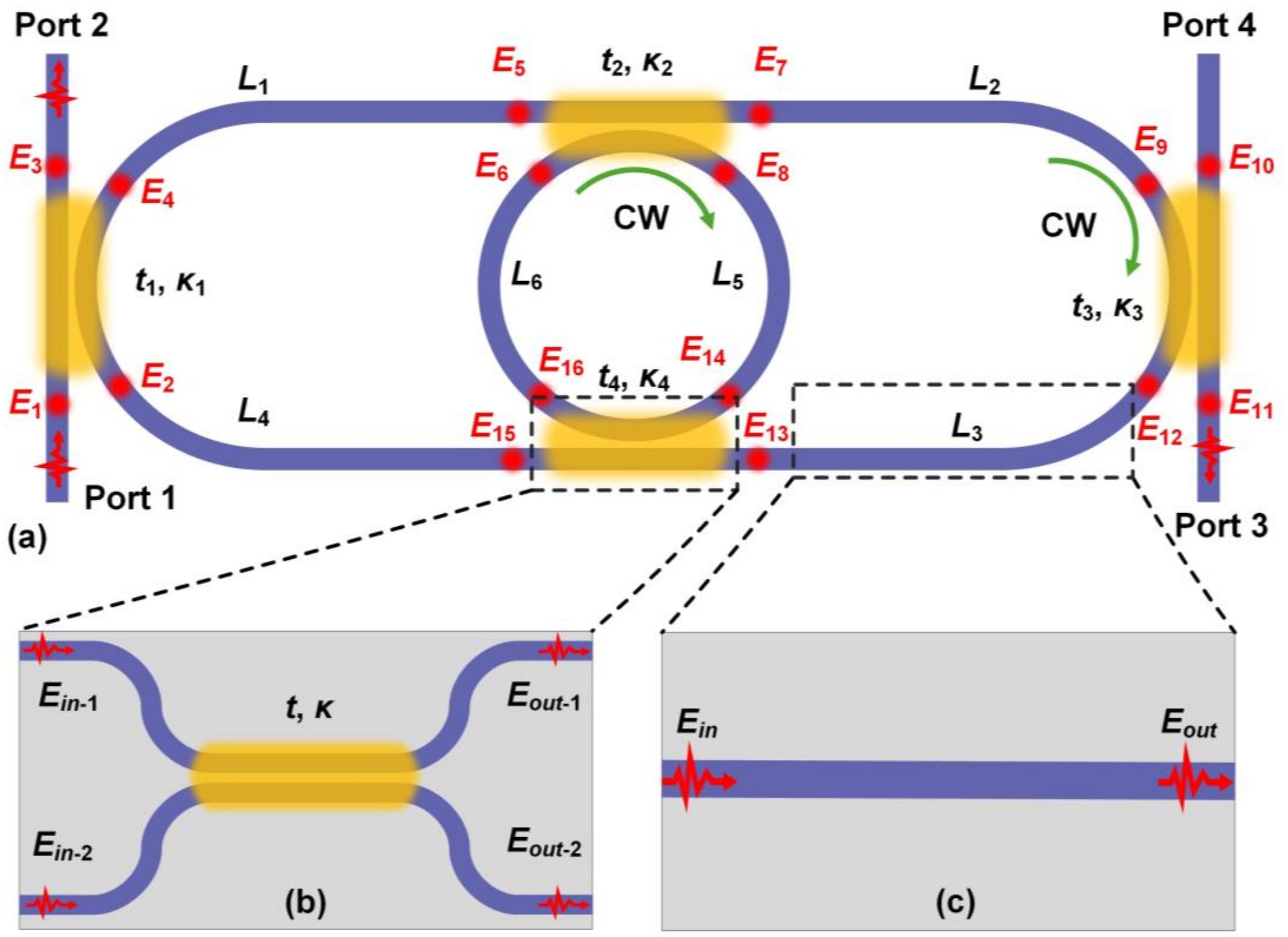

**Figure 1.** (a) Schematic illustration of an integrated photonic resonator consisting of two microring resonators (MRRs) with an input from Port 1. The device is divided into several directional couplers and connecting waveguides, with $E_i$ ($i$ = 1–16) denoting the optical fields at the dividing points. (b) – (c) zoom-in views for a directional coupler and a connecting waveguide, respectively.

Next, we establish a set of scattering matrix equations for the basic elements obtained in the first step, as shown in **Table 1**. These scattering matrix equations describe the relation between $E_i$ at the input / output ports of the directional couplers and the connecting waveguides. For the directional coupler in **Figure 1(b)**, the field transfer function can be expressed as [16]

$$\begin{bmatrix} E_{out\text{-}1} \\ E_{out\text{-}2} \end{bmatrix} = \begin{bmatrix} t & j\kappa \\ j\kappa & t \end{bmatrix} \begin{bmatrix} E_{in\text{-}1} \\ E_{in\text{-}2} \end{bmatrix}, \quad (1)$$

where $j = \sqrt{-1}$, $t$ and $\kappa$ are the self-coupling and cross coupling coefficients, which satisfies the relation $t^2 + \kappa^2 = 1$ when assuming lossless coupling, $E_{in\text{-}1}$, $E_{in\text{-}2}$, $E_{out\text{-}1}$, and $E_{out\text{-}2}$ are the input and output optical fields right before and after the coupling region. For the connecting waveguide in **Figure 1(c)**, the field transfer function can be given by [23]

$$T = ae^{-j\varphi}, \quad (2)$$

where $a = e^{-\alpha L/2}$ is the round-trip transmission factor, with $\alpha$ and $L$ denoting the power propagation loss factor and the waveguide length, respectively. In **Eq. (2)**, $\varphi = 2\pi n_g L/\lambda$ is the round-trip phase shift, with $n_g$ and $\lambda$ denoting the group index and the wavelength, respectively. For each directional coupler, two equations can be derived, and one equation can be obtained for each connecting waveguide. This results in $2 \times 4 + 1 \times 6 = 14$ equations in total for the device shown in **Figure 1(a)**, which includes four directional couplers and six connecting waveguides.

**Table 1. Definitions of structural parameters of the device in Figure 1(a) and the corresponding scattering matrix equations.**

| | | | | |
|---|---|---|---|---|
| **Directional couplers** | Structural Parameters ($i$ = 1 – 4) | Field transmission coefficient $t_i$ | Field cross coupling coefficient $\kappa_i$ | |
| **Directional couplers** | Scattering matrix equations | $\begin{bmatrix} E_3 \\ E_4 \end{bmatrix} = \begin{bmatrix} t_1 & j\kappa_1 \\ j\kappa_1 & t_1 \end{bmatrix} \begin{bmatrix} E_1 \\ E_2 \end{bmatrix}$, $\begin{bmatrix} E_7 \\ E_8 \end{bmatrix} = \begin{bmatrix} t_2 & j\kappa_2 \\ j\kappa_2 & t_2 \end{bmatrix} \begin{bmatrix} E_5 \\ E_6 \end{bmatrix}$, $\begin{bmatrix} E_{12} \\ E_{11} \end{bmatrix} = \begin{bmatrix} t_3 & j\kappa_3 \\ j\kappa_3 & t_3 \end{bmatrix} \begin{bmatrix} E_9 \\ E_{10} \end{bmatrix}$, $\begin{bmatrix} E_{15} \\ E_{16} \end{bmatrix} = \begin{bmatrix} t_4 & j\kappa_4 \\ j\kappa_4 & t_4 \end{bmatrix} \begin{bmatrix} E_{13} \\ E_{14} \end{bmatrix}$. | | |
| **Connecting waveguides** | Structural Parameters ($i$ = 1 – 6) | Length $L_i$ | Transmission factor $a_i$ | Phase shift $\varphi_i$ |
| **Connecting waveguides** | Scattering matrix equations | $E_5 = T_1E_4$, $E_9 = T_2E_7$, $E_{13} = T_3E_{12}$, $E_2 = T_4E_{15}$, $E_{14} = T_5E_8$, $E_6 = T_6E_{16}$. | | |
| **Input** | | $E_1 = 1$, $E_{10} = 0$. | | |

Third, the system input is set. For instance, if we assume that there is only a continuous-wave (CW) input from Port 1, then another two equations can be obtained: $E_1 = 1$ and $E_{10} = 0$. Here we set $E_1$ as 1 because the spectral transfer function at the output port, such as Port 2, is given by $f_{\text{Port 2}} = E_3 / E_1$. By setting $E_1$ to 1, $f_{\text{Port 2}} = E_3 / E_1 = E_3$, then the transfer function can be directly determined by calculating $E_3$ in the next step.

Finally, by solving all the linear equations obtained in the second and third steps, one can obtain the spectral transfer functions at the output ports. In **Table 1**, we summarize all the 16 equations for the device in **Figure 1(a)**. In these equations, there are 16 variables $E_i$ ($i$ = 1–16), with the device's structural parameters denoted using symbols and treated as constant coefficients. By solving this system of linear equations using computational tools (*e.g.*, symbolic calculation in MATLAB), any of the variables $E_i$ ($i$ = 1–16) can be determined as a function of the device's structural parameters. For example, the spectral transfer function at Port 2 can be given by $f_{\text{Port 2}} = E_3 (T_i, t_i, \kappa_i)$. Assuming that the power propagation loss factor $\alpha$ and the group index $n_g$ are constant for the device, the spectral transfer function at Port 2 can be expressed as a function of $L_i$, $t_i$, and $\kappa_i$, *i.e.*, $f_{\text{Port 2}} = E_3 (L_i, t_i, \kappa_i)$.

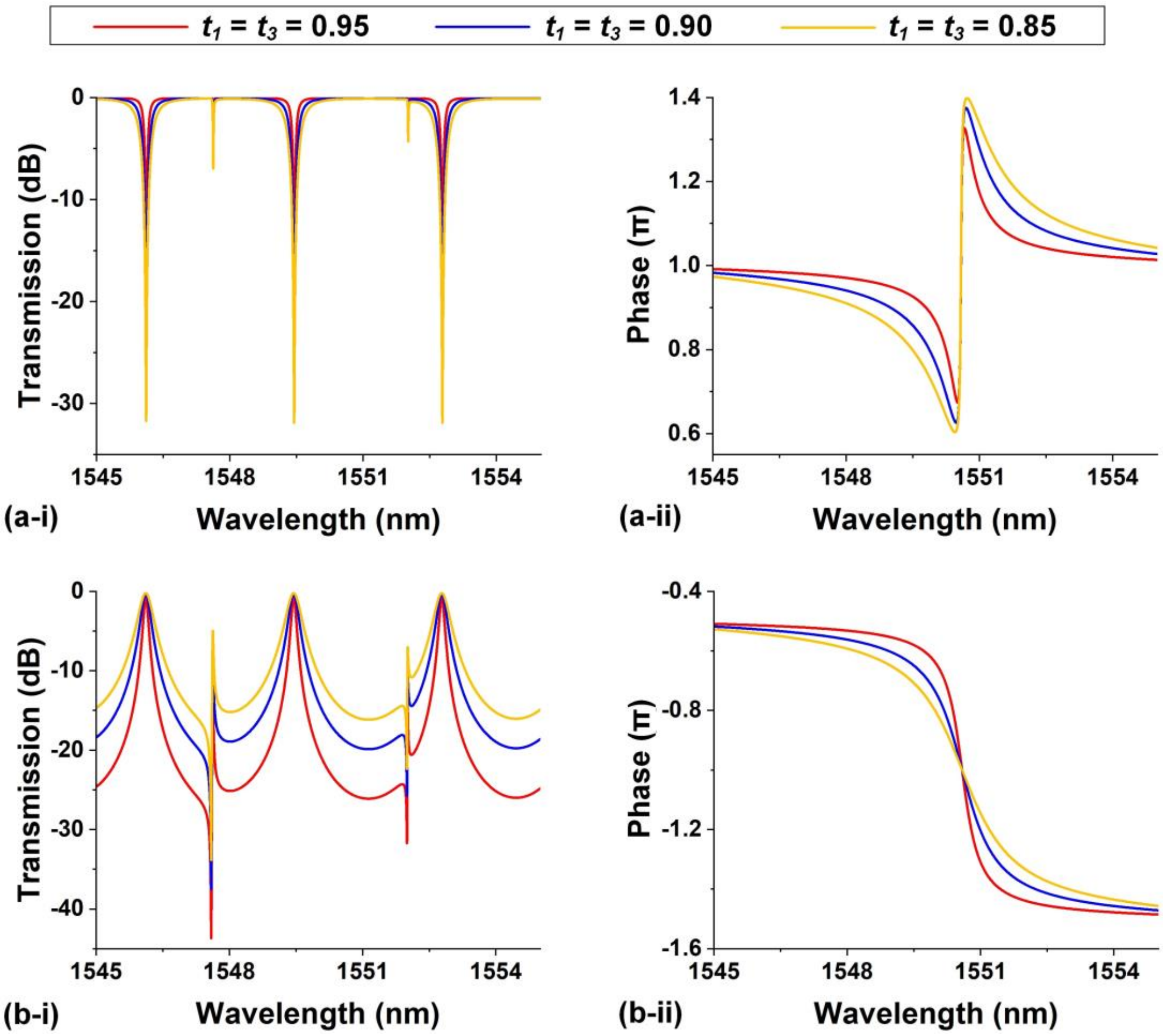

**Figure 2.** (a) Calculated (i) intensity and (ii) phase response spectra at Port 2 for the device in **Figure 1** with different $t_1 = t_3$. (b) Calculated (i) intensity and (ii) phase response spectra at Port 3 for the device in **Figure 1** with different $t_1 = t_3$. In (a) and (b), $t_2 = t_4 = 0.98$. The power propagation loss factor $\alpha$ and group index $n_g$ are assumed to be 101 m$^{-1}$ and 4.335, respectively.

In **Figure 2(a)**, we show the intensity and phase response spectra at Port 2 for different $t_i$ based on the obtained spectral transfer function $f_{\text{Port 2}} = E_2\,(L_i,\ t_i,\ \kappa_i)$. Except for the varying parameters, other structural parameters were chosen as follows: $L_1 = L_2 = L_3 = L_4 =$ 41.73 μm, $L_5 = L_6 =$ 62.83 μm, and $t_2 = t_4 = 0.98$. By using the same method, we obtain the intensity and phase responses at Port 3 for different $t_i$, as shown in **Figure 2(b)**. In our modeling, we assume that $\alpha = 101$ m$^{-1}$ and group index $n_g = 4.335$ based on values obtained from our previously fabricated silicon photonic devices [20, 25]. Unless otherwise specified, we use the same values for these two parameters in the modeling for all the devices in the following figures.

Our results in **Figure 2** show good agreement with the results in Ref. [26], confirming the effectiveness of our method. Since the obtained spectral transfer functions (*e.g.*, $f_{\text{Port 2}} = E_2\,(L_i,\ t_i,\ \kappa_i)$) include all the structural parameters, the intensity and phase responses can be easily tailored by adjusting any of these parameters. This provides significant flexibility in designing and optimizing the spectral response of integrated photonic devices with different structural parameters. It should also be noted that our modeling, as mentioned above, is not limited to silicon photonic platforms. By using the corresponding values of $\alpha$ and $n_g$, it is applicable across various integrated platforms such as silicon nitride, doped silica, and lithium niobate [27-93].

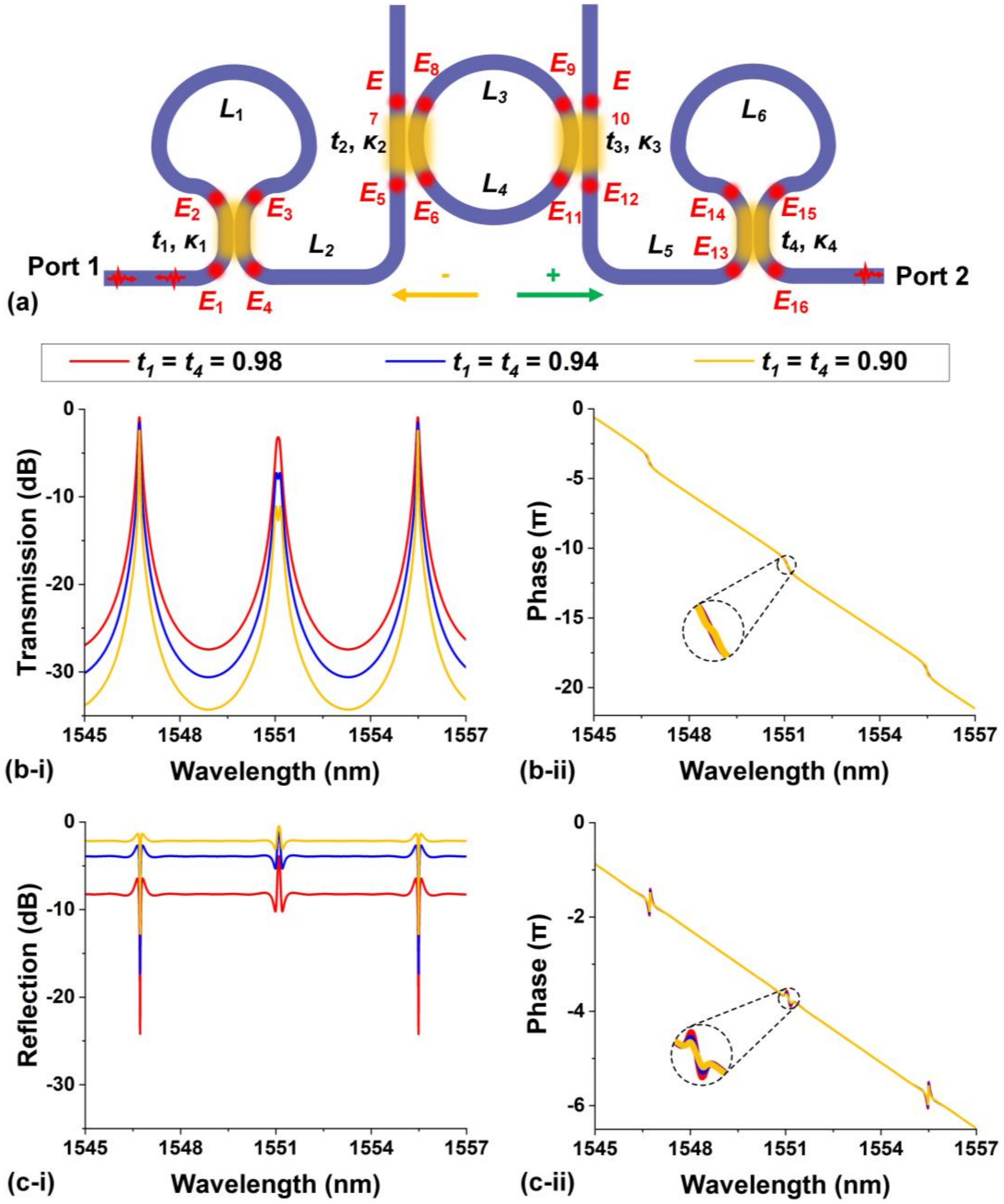

**Figure 3.** (a) Schematic illustration of an integrated photonic resonator formed by an add-drop micro-ring resonator (AD-MRR) sandwiched between two Sagnac interferometers (SIs) with an input from Port 1. The device is divided into several directional couplers and connecting waveguides, with $E_i$ ($i$ = 1–16) denoting the optical fields at the dividing points. The electric fields propagating from left to right are defined as "+" direction and the ones propagating from right to left are defined as "–" direction. (b) Calculated (i) power transmission and (ii) phase response spectra at Port 2 for different $t_1 = t_4$. (c) Calculated (i) power reflection and (ii) phase response spectra at Port 1 for different $t_1 = t_4$.

## 3. Modeling of devices with bidirectional light propagation

For the device in **Figure 1(a)**, light propagates in only one direction along each waveguide. In more complex integrated photonic resonators with light propagating bidirectionally in the waveguides, the modeling becomes more complicated. In this section, we use the device configuration shown in **Figure 3(a)** as an example to show how to model a complex integrated photonic resonator with bidirectional light propagation based on the SMM. The device consists of an add-drop (AD) MRR sandwiched between a pair of Sagnac interferometers (SIs), where the SIs introduce back reflection and hence bidirectional light propagation in such device.

To model the device in Figu**re 3(a)**, we first divide it into several directional couplers and connecting waveguides – the same as what we did for the device in **Figure 1(a)**. The optical fields at the dividing points between directional couplers and connecting waveguides are denoted as $E_i$ ($i$ = 1–16). Since there are optical fields traveling in two different directions at each dividing point, the optical fields traveling from left to right or in a clockwise direction are defined as "+", and the opposite direction is defined as "–".

In the second step, we establish 28 scattering matrix equations describing the relation between $E_i^+$ and $E_i^-$ at the input / output ports of the directional couplers and the connecting waveguides, as shown in **Table 2**. For each directional coupler, four equations can be derived, and two equations can be obtained for each connecting waveguide. Note that the number of equations is doubled due to the bidirectional light propagation in such device.

In the third step, we obtain another four equations by setting the system input. If we assume that there is only a CW input from Port 1, then the four equations are: $E_1^+ = 1$, $E_7^- = 0$, $E_{10}^+ = 0$, and $E_{16}^- = 0$. Here the number of equations is also doubled due to the bidirectional light propagation.

Finally, by solving all the 32 linear equations obtained previously (which include 32 variables, *i.e.*, $E_i^+$ and $E_i^-$, $i$ = 1–16), we can obtain the spectral transfer functions at all the output ports. For example, the transfer function for forward transmission, with output at Port 2, is $f_{\text{Port 2}} = E_{16}^+ (L_i, t_i, \kappa_i)$, whereas the transfer function for backward reflection is $f_{\text{Port 1}} = E_1^- (L_i, t_i, \kappa_i)$, with output at Port 1. It should be noted that as the number of equations increases or the device configuration becomes more complex, the use of computational software to solve the system of linear equations offers significant advantages as compared to manual derivation of the spectral transfer functions as those in Refs. [25, 26].

**Table 2. Definitions of structural parameters of the device in Figure 3(a) and the corresponding scattering matrix equations.**

| | | | | |
|---|---|---|---|---|
| **Directional couplers** | Structural Parameters ($i$ = 1 – 4) | Field transmission coefficient $t_i$ | | Field cross coupling coefficient $\kappa_i$ |
| | Scattering matrix equations | $\begin{bmatrix} E_2^+ \\ E_3^- \end{bmatrix} = \begin{bmatrix} t_1 & j\kappa_1 \\ j\kappa_1 & t_1 \end{bmatrix} \begin{bmatrix} E_1^+ \\ E_4^- \end{bmatrix}, \begin{bmatrix} E_1^- \\ E_4^+ \end{bmatrix} = \begin{bmatrix} t_1 & j\kappa_1 \\ j\kappa_1 & t_1 \end{bmatrix} \begin{bmatrix} E_2^- \\ E_3^+ \end{bmatrix},$ $\begin{bmatrix} E_7^+ \\ E_8^+ \end{bmatrix} = \begin{bmatrix} t_2 & j\kappa_2 \\ j\kappa_2 & t_2 \end{bmatrix} \begin{bmatrix} E_5^+ \\ E_6^+ \end{bmatrix}, \begin{bmatrix} E_5^- \\ E_6^- \end{bmatrix} = \begin{bmatrix} t_2 & j\kappa_2 \\ j\kappa_2 & t_2 \end{bmatrix} \begin{bmatrix} E_7^- \\ E_8^- \end{bmatrix},$ $\begin{bmatrix} E_{11}^+ \\ E_{12}^+ \end{bmatrix} = \begin{bmatrix} t_3 & j\kappa_3 \\ j\kappa_3 & t_3 \end{bmatrix} \begin{bmatrix} E_9^+ \\ E_{10}^+ \end{bmatrix}, \begin{bmatrix} E_9^- \\ E_{10}^- \end{bmatrix} = \begin{bmatrix} t_3 & j\kappa_3 \\ j\kappa_3 & t_3 \end{bmatrix} \begin{bmatrix} E_{11}^- \\ E_{12}^- \end{bmatrix},$ $\begin{bmatrix} E_{14}^+ \\ E_{15}^- \end{bmatrix} = \begin{bmatrix} t_4 & j\kappa_4 \\ j\kappa_4 & t_4 \end{bmatrix} \begin{bmatrix} E_{13}^+ \\ E_{16}^- \end{bmatrix}, \begin{bmatrix} E_{13}^- \\ E_{16}^+ \end{bmatrix} = \begin{bmatrix} t_4 & j\kappa_4 \\ j\kappa_4 & t_4 \end{bmatrix} \begin{bmatrix} E_{14}^- \\ E_{15}^+ \end{bmatrix}.$ | | |
| **Connecting waveguides** | Structural Parameters ($i$ = 1 – 6) | Length $L_i$ | Transmission factor $a_i$ | Phase shift $\varphi_i$ |
| | Scattering matrix equations | $\begin{bmatrix} E_3^+ \\ E_2^- \end{bmatrix} = T_1 \begin{bmatrix} E_2^+ \\ E_3^- \end{bmatrix}, \begin{bmatrix} E_5^+ \\ E_4^- \end{bmatrix} = T_2 \begin{bmatrix} E_4^+ \\ E_5^- \end{bmatrix}, \begin{bmatrix} E_9^+ \\ E_8^- \end{bmatrix} = T_3 \begin{bmatrix} E_8^+ \\ E_9^- \end{bmatrix},$ $\begin{bmatrix} E_6^+ \\ E_{11}^- \end{bmatrix} = T_4 \begin{bmatrix} E_{11}^+ \\ E_6^- \end{bmatrix}, \begin{bmatrix} E_{13}^+ \\ E_{12}^- \end{bmatrix} = T_5 \begin{bmatrix} E_{12}^+ \\ E_{13}^- \end{bmatrix}, \begin{bmatrix} E_{15}^+ \\ E_{14}^- \end{bmatrix} = T_6 \begin{bmatrix} E_{14}^+ \\ E_{15}^- \end{bmatrix}.$ | | |
| **Input** | $E_1^+ = 1$, $E_7^- = 0$, $E_{10}^+ = 0$, $E_{16}^- = 0$. | | | |

Based on the obtained spectral transfer functions at Port 2 and Port 1, we plot the corresponding response spectra for different $t_i$, as shown in **Figures 3(b)** and **3(c),** respectively. Except for $t_i$, the other structural parameters are kept as constant as: $L_1 = L_6$ = 129.66 μm, $L_2 = L_5$ = 77.67 μm, $L_3 = L_4$ = 63.33 μm, and $t_2 = t_3$ = 0.95. Our results in **Figure 3** also show good agreement with the results in Ref. [25], confirming the effectiveness of our method in modeling complex integrated photonic resonator with bidirectional light propagation.

**4. Simplification by dividing device into submodules**

In modeling of the device configurations in **Figures 1(a)** and **3(a)**, the devices were divided into several basic elements including directional couplers and connecting waveguides. To reduce the number of equations and simplify the calculation of the spectral transfer function, the device configuration can also be divided into several independent submodules such as MRRs and SIs, which are formed by the basic elements.

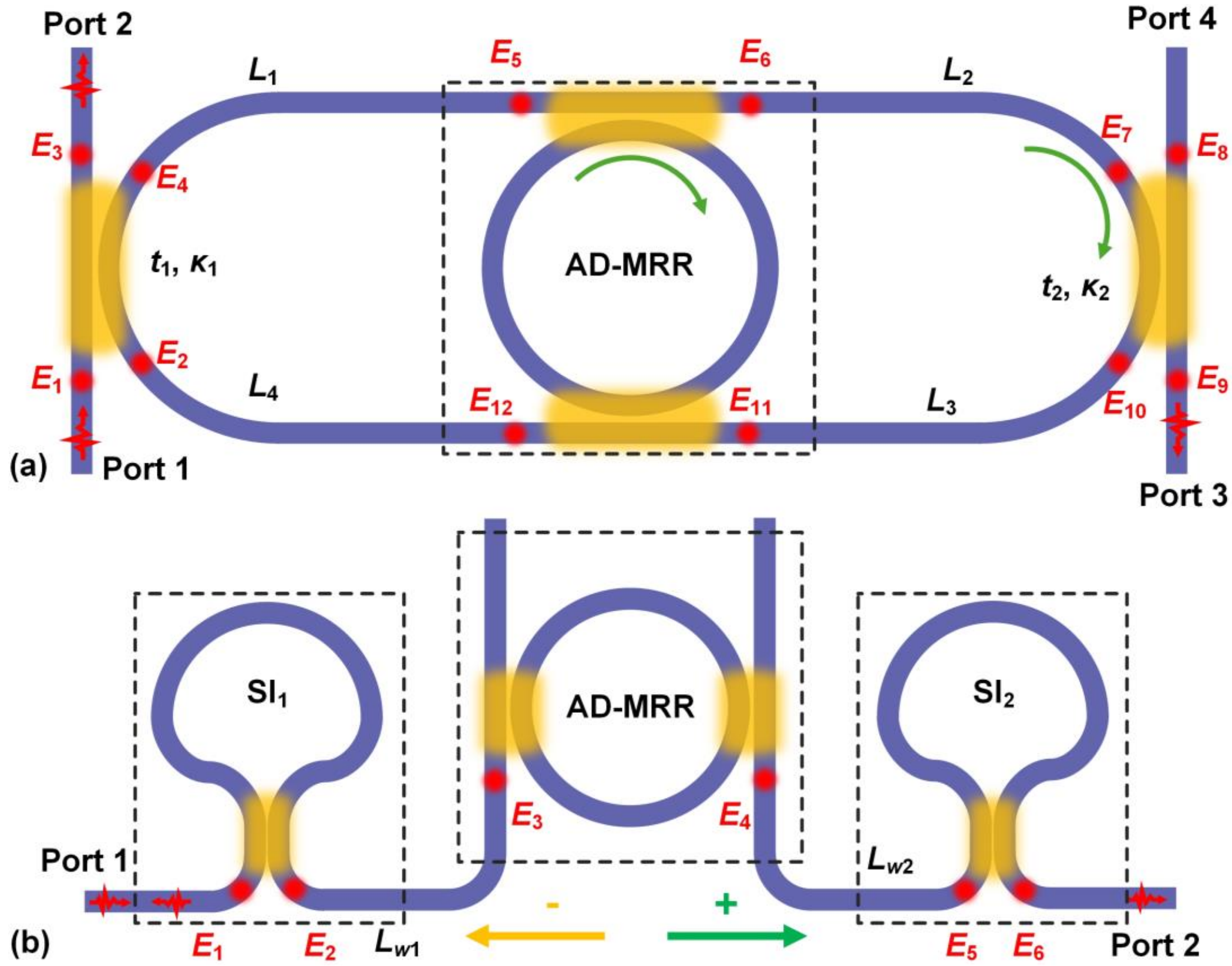

**Figure 4.** Dividing device configuration into submodules to simplify the calculation (a) Device configuration in **Figure 1(a)**, where the add-drop micro-ring resonator (AD-MRR) at the center is regarded as an independent module. (b) Device configuration in **Figure 3(a)**, which is divided into submodules including an AD-MRR, two Sagnac interferometers (SIs), and connecting waveguides between them.

As illustrated in **Figure 4(a)**, the AD-MRR at the center of the device in **Figure 1(a)** is treated as an independent module, with the field transfer functions at the through and drop ports denoted as $T_{RR}$ and $D_{RR}$, respectively. This results in 12 equations in total obtained by using our method mentioned previously, as summarized in **Table 3**. In comparison to the 16 equations obtained for the same device configuration in **Table 1**, the number of equations is reduced, which helps to simplify the solving of the system of equations. By solving the 12 equations in **Table 3**, the spectral transfer function at Port 2 can be obtained, which is a function of $T_{MRR}$ and $D_{MRR}$. We compared the response spectra obtained by solving the equations in **Tables 1** and **3** and found them to be identical. This verifies the validity of the new dividing method in **Figures 4(a)**. It is also worth mentioning that when plotting the response spectra based on this dividing method, $T_{MRR}$ and $D_{MRR}$ should be further expressed as [94]

$$T_{RR} = \frac{E_6}{E_5} = \frac{E_{12}}{E_{11}} = \frac{t_{RR\text{-}1} - t_{RR\text{-}2}a_{RR}e^{-j\varphi_{RR}}}{1 - t_{RR\text{-}1}t_{RR\text{-}2}a_{RR}e^{-j\varphi_{RR}}}, \quad (3)$$

$$D_{RR} = \frac{E_{12}}{E_5} = \frac{E_6}{E_{11}} = \frac{-\kappa_{RR\text{-}1}\kappa_{RR\text{-}2}\sqrt{a_{RR}}e^{-j\varphi_{RR}}}{1 - t_{RR\text{-}1}t_{RR\text{-}2}a_{RR}e^{-j\varphi_{RR}}}, \quad (4)$$

where $j = \sqrt{-1}$, $a_{RR}$ and $\varphi_{RR}$ are the round-trip transmission factor and phase shift along the ring in the AD-MRR, respectively. $t_{RR\text{-}i}$ and $\kappa_{RR\text{-}i}$ ($i$ = 1, 2) are the self-coupling and cross coupling coefficients of the directional couplers in the AD-MRR, respectively.

**Table 3. Scattering matrix equations for the device in Figure 4(a). $T_{RR}$ and $D_{RR}$ are the field transfer functions at the through and drop ports for the AD-MRR, respectively.**

| | | |
|---|---|---|
| **Scattering matrix equations** | Add-drop micro-ring resonator (AD-MRR) | $\begin{bmatrix} E_6 \\ E_{12} \end{bmatrix} = \begin{bmatrix} T_{RR} & D_{RR} \\ D_{RR} & T_{RR} \end{bmatrix} \begin{bmatrix} E_5 \\ E_{11} \end{bmatrix}$. |
| | Directional couplers | $\begin{bmatrix} E_3 \\ E_4 \end{bmatrix} = \begin{bmatrix} t_1 & j\kappa_1 \\ j\kappa_1 & t_1 \end{bmatrix} \begin{bmatrix} E_1 \\ E_2 \end{bmatrix}$, $\begin{bmatrix} E_{10} \\ E_9 \end{bmatrix} = \begin{bmatrix} t_2 & j\kappa_2 \\ j\kappa_2 & t_2 \end{bmatrix} \begin{bmatrix} E_7 \\ E_8 \end{bmatrix}$. |
| | Connecting waveguides | $E_5 = T_{w1}E_4$, $E_7 = T_{w2}E_6$, $E_{11} = T_{w3}E_{10}$, $E_2 = T_{w4}E_{12}$. |
| **Input** | | $E_1 = 1$, $E_8 = 0$. |

Similarly, in **Figure 4(b)** we divided the device configuration in **Figure 3(a)** into an add-drop MRR, two SIs, and several connecting waveguides between them. By using this new dividing method, we obtained the same response spectra as those in **Figures 3(b)** and **3(c).** When plotting the response spectra, the forward and backward field transfer functions for the SIs were further expressed as

$$T_{SI\text{-}i} = \frac{E_2^+}{E_1^+} = \frac{E_6^+}{E_5^+} = (t_{SI\text{-}i}^2 - \kappa_{SI\text{-}i}^2)a_{SI\text{-}i}e^{-j\varphi_{SI\text{-}i}}, \quad (i = 1, 2) \quad (5)$$

$$R_{SI\text{-}i} = \frac{E_1^-}{E_1^+} = \frac{E_5^-}{E_5^+} = 2jt_{SI\text{-}i}\kappa_{SI\text{-}i}a_{SI\text{-}i}e^{-j\varphi_{SI\text{-}i}}, \quad (i = 1, 2) \quad (6)$$

where $j = \sqrt{-1}$, $a_{SI\text{-}i}$ and $\varphi_{SI\text{-}i}$ ($i$ = 1, 2) are the round-trip transmission factor and phase shift along rings in the SIs, respectively. $t_{SI\text{-}i}$ and $\kappa_{SI\text{-}i}$ ($i$ = 1, 2) are the self-coupling and cross coupling coefficients of the directional couplers in the SIs, respectively.

As shown in **Table 4**, only 10 equations are established by using the new dividing method, in contrast to 32 equations obtained for the dividing method in **Figure 3(a)**. This indicates that although dividing device into basic elements is a universal approach, dividing it into submodules simplifies the process of solving equations. This is particularly true for devices composed of multiple submodules, where the advantages of the new division method become even more evident.

**Table 4. Scattering matrix equations for the device in Figure 4(b). $T_{SI\text{-}i}$ and $R_{SI\text{-}i}$ ($i$ = 1, 2) are forward and backward field transfer functions for the SIs, respectively. $T_{RR}$ and $D_{RR}$ are the field transfer functions at the through and drop ports for the AD-MRR, respectively.**

| | | |
|---|---|---|
| **Scattering matrix equations** | Sagnac interferometers (SIs) | $\begin{bmatrix} E_2^+ \\ E_1^- \end{bmatrix} = \begin{bmatrix} T_{SI\text{-}1} & R_{SI\text{-}1} \\ R_{SI\text{-}1} & T_{SI\text{-}1} \end{bmatrix} \begin{bmatrix} E_1^+ \\ E_2^- \end{bmatrix}$, $\begin{bmatrix} E_6^+ \\ E_5^- \end{bmatrix} = \begin{bmatrix} T_{SI\text{-}2} & R_{SI\text{-}2} \\ R_{SI\text{-}2} & T_{SI\text{-}2} \end{bmatrix} \begin{bmatrix} E_5^+ \\ E_6^- \end{bmatrix}$. |
| | Add-drop micro-ring resonator (AD-MRR) | $\begin{bmatrix} E_4^+ \\ E_3^- \end{bmatrix} = D_{RR} \begin{bmatrix} E_3^+ \\ E_4^- \end{bmatrix}$. |
| | Connecting waveguides | $\begin{bmatrix} E_3^+ \\ E_2^- \end{bmatrix} = T_{w1} \begin{bmatrix} E_2^+ \\ E_3^- \end{bmatrix}$, $\begin{bmatrix} E_5^+ \\ E_4^- \end{bmatrix} = T_{w2} \begin{bmatrix} E_4^+ \\ E_5^- \end{bmatrix}$. |
| **Input** | | $E_1^+ = 1$, $E_6^- = 0$. |

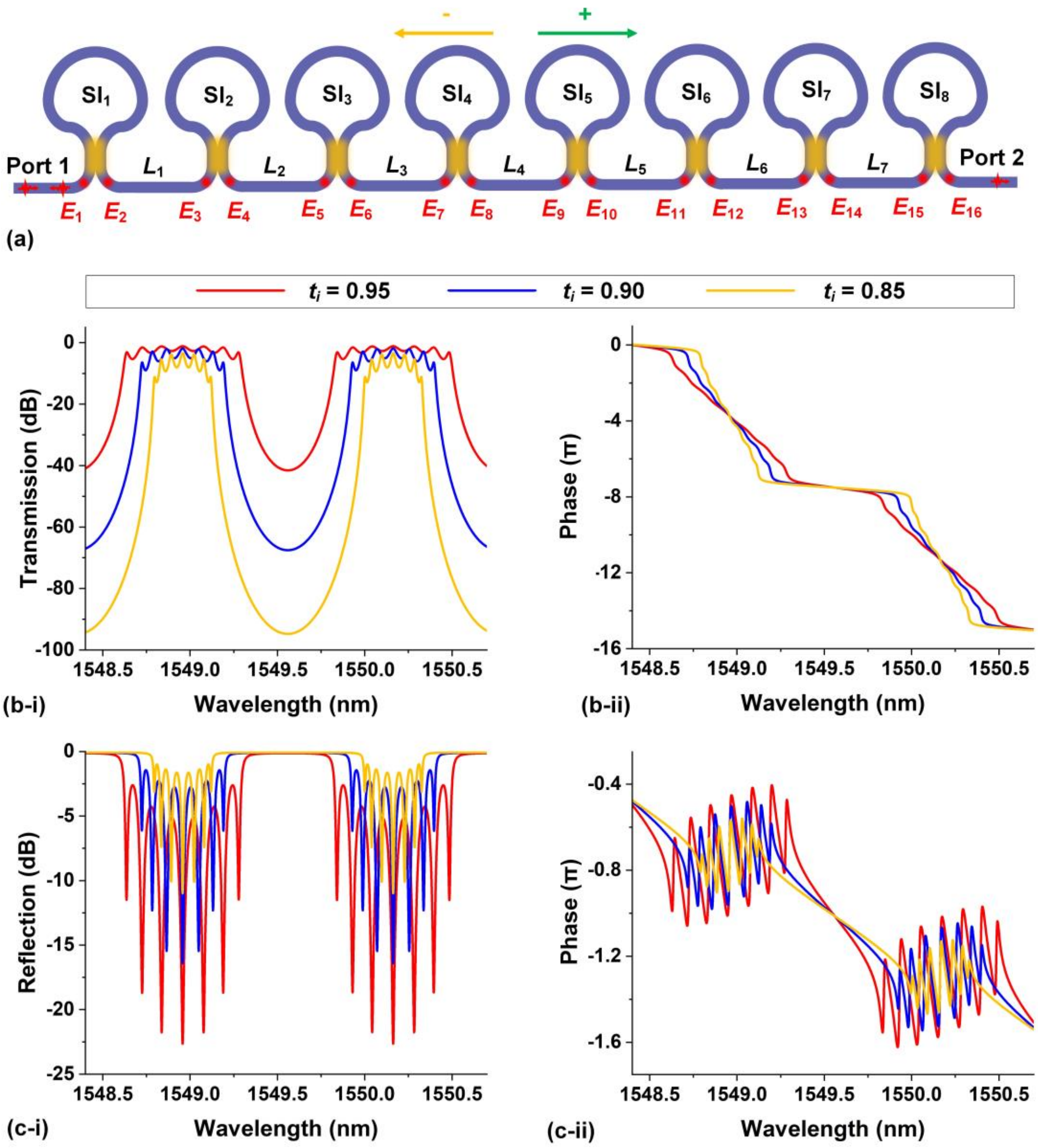

**Figure 5.** (a) Schematic illustration of an integrated photonic resonator formed by eight cascaded Sagnac interferometers (SIs) with input from Port 1. The device is divided into eight SIs and connecting waveguides between them, with $E_i$ ($i$ = 1–16) denoting the optical fields at the dividing points. The electric fields propagating from to right to left are defined as "+" direction and the ones propagating from left to right are defined as "–" direction. (b) Calculated (i) power transmission and (ii) phase response spectra at Port 2 for different $t_i$ ($i$ = 1–8), which are the are the self-coupling coefficients of the directional couplers in SIs. (c) Calculated (i) power reflection and (ii) phase response spectra at Port 1 for different $t_i$ ($i$ = 1–8). In (b) and (c), the circumference of the rings in the SIs are 129.66 μm, and the lengths of the connecting waveguides between them are 100 μm.

In **Figure 5**, we model a more complex device by dividing it into eight SIs and establishing 32 linear equations, as summarized in **Table 5**. The forward and backward field transfer functions for the SIs are denoted as $T_{SI\text{-}i}$ and $R_{SI\text{-}i}$ ($i = 1 - 8$), respectively. Dividing the device into basic elements would result in a system of 64 equations, compared to the 32 equations shown in **Table 5**. By solving the system of 32 equations to obtain $E_{16}^{+}$ and $E_1^{-}$, we obtain the spectral transfer functions at Port 2 and Port 1 (reflection), respectively. The corresponding response spectra are plotted in **Figures 5(b)** and **5(c)**, which are consistent with the results in Ref. [95] and further confirms the effectiveness of our method in modeling devices with complex configurations.

**Table 5. Scattering matrix equations for the device in Figure 5(a). $T_{SI\text{-}i}$ and $R_{SI\text{-}i}$ ($i$ = 1 – 8) are the forward and backward field transfer functions for the SIs, respectively.**

| | | |
|---|---|---|
| **Scattering matrix equations** | Sagnac interferometers (SIs) | $\begin{bmatrix}E_2^+\\E_1^-\end{bmatrix} = \begin{bmatrix}T_{SI\text{-}1} & R_{SI\text{-}1}\\R_{SI\text{-}1} & T_{SI\text{-}1}\end{bmatrix}\begin{bmatrix}E_1^+\\E_2^-\end{bmatrix}, \begin{bmatrix}E_4^+\\E_3^-\end{bmatrix} = \begin{bmatrix}T_{SI\text{-}2} & R_{SI\text{-}2}\\R_{SI\text{-}2} & T_{SI\text{-}2}\end{bmatrix},$ $\begin{bmatrix}E_6^+\\E_5^-\end{bmatrix} = \begin{bmatrix}T_{SI\text{-}3} & R_{SI\text{-}3}\\R_{SI\text{-}3} & T_{SI\text{-}3}\end{bmatrix}\begin{bmatrix}E_5^+\\E_6^-\end{bmatrix}, \begin{bmatrix}E_8^+\\E_7^-\end{bmatrix} = \begin{bmatrix}T_{SI\text{-}4} & R_{SI\text{-}4}\\R_{SI\text{-}4} & T_{SI\text{-}4}\end{bmatrix}\begin{bmatrix}E_7^+\\E_8^-\end{bmatrix},$ $\begin{bmatrix}E_{10}^+\\E_9^-\end{bmatrix} = \begin{bmatrix}T_{SI\text{-}5} & R_{SI\text{-}5}\\R_{SI\text{-}5} & T_{SI\text{-}5}\end{bmatrix}\begin{bmatrix}E_9^+\\E_{10}^-\end{bmatrix}, \begin{bmatrix}E_{12}^+\\E_{11}^-\end{bmatrix} = \begin{bmatrix}T_{SI\text{-}6} & R_{SI\text{-}6}\\R_{SI\text{-}6} & T_{SI\text{-}6}\end{bmatrix}\begin{bmatrix}E_{11}^+\\E_{12}^-\end{bmatrix},$ $\begin{bmatrix}E_{14}^+\\E_{13}^-\end{bmatrix} = \begin{bmatrix}T_{SI\text{-}7} & R_{SI\text{-}7}\\R_{SI\text{-}7} & T_{SI\text{-}7}\end{bmatrix}\begin{bmatrix}E_{13}^+\\E_{14}^-\end{bmatrix}, \begin{bmatrix}E_{16}^+\\E_{15}^-\end{bmatrix} = \begin{bmatrix}T_{SI\text{-}8} & R_{SI\text{-}8}\\R_{SI\text{-}8} & T_{SI\text{-}8}\end{bmatrix}\begin{bmatrix}E_{15}^+\\E_{16}^-\end{bmatrix}.$ |
| | Connecting waveguides | $\begin{bmatrix}E_3^+\\E_2^-\end{bmatrix} = T_{w1}\begin{bmatrix}E_2^+\\E_3^-\end{bmatrix}, \begin{bmatrix}E_5^+\\E_4^-\end{bmatrix} = T_{w2}\begin{bmatrix}E_4^+\\E_5^-\end{bmatrix}, \begin{bmatrix}E_7^+\\E_6^-\end{bmatrix} = T_{w3}\begin{bmatrix}E_6^+\\E_7^-\end{bmatrix},$ $\begin{bmatrix}E_9^+\\E_8^-\end{bmatrix} = T_{w4}\begin{bmatrix}E_8^+\\E_9^-\end{bmatrix}, \begin{bmatrix}E_{11}^+\\E_{10}^-\end{bmatrix} = T_{w5}\begin{bmatrix}E_{10}^+\\E_{11}^-\end{bmatrix}, \begin{bmatrix}E_{13}^+\\E_{12}^-\end{bmatrix} = T_{w6}\begin{bmatrix}E_{12}^+\\E_{13}^-\end{bmatrix},$ $\begin{bmatrix}E_{15}^+\\E_{14}^-\end{bmatrix} = T_{w7}\begin{bmatrix}E_{14}^+\\E_{15}^-\end{bmatrix}.$ |
| **Input** | | $E_{16}^+ = 1$, $E_1^- = 0$. |

When dividing complex devices into submodules, it's important to ensure that the submodules are independent. Here, "independent" means that the submodules exchange energy with other parts exclusively through connecting waveguides. For the device shown in **Figure 6a** formed by a self-coupled waveguide, even though there are multiple SIs within the device, it cannot be divided into submodules of SIs. This is because the adjacent SIs are mutually coupled with energy exchange through directional couplers. In this case, the spectral transfer functions can be obtained by using the universal dividing method, which divides the device into basic elements of directional couplers and connecting waveguides.

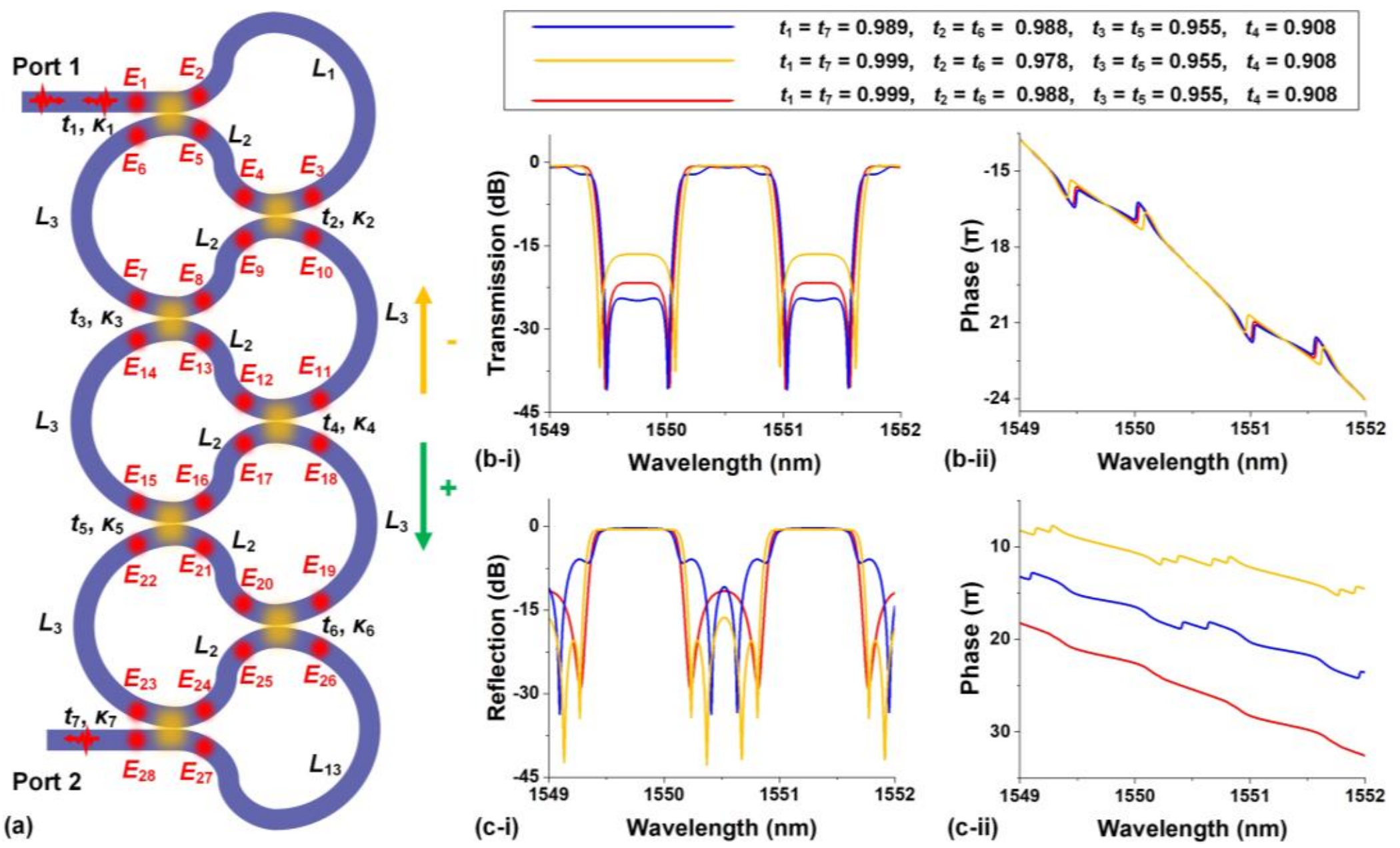

**Figure 6.** (a) Schematic illustration of an integrated photonic resonator formed by a self-coupled waveguide with input from Port 1. The device is divided into several directional couplers and connecting waveguides, with $E_i$ ($i$ = 1–28) denoting the optical fields at the dividing points. The electric fields propagating from to up to down are defined as "+" direction and the ones propagating from down to up are defined as "−" direction. (b) Calculated (i) power transmission and (ii) phase response spectra at Port 2 for different $t_i$, which are the are the self-coupling coefficients of the directional couplers. (c) Calculated (i) power reflection and (ii) phase response spectra at Port 1 for different $t_i$.

**Table 6** shows a system of 56 equations obtained by using the universal dividing method for the device in **Figure 6a**. By solving these equations to obtain $E_{28}^{+}$ and $E_{1}^{-}$, we obtain the spectral transfer functions at Port 1 and Port 2, respectively. The corresponding response spectra for different $t_i$ are shown in **Figures 6(b)** and **6(c)**. Except for $t_i$, all other structural parameters are kept constant as: $L_1$ = 180 μm, $L_2$ = 60 μm and $L_3$ = 120 μm. The response spectra also show a good agreement with the results in Ref. [24], providing additional evidence of the effectiveness of our method.

**Table 6. Definitions of structural parameters of the device in Figure 6(a) and the corresponding scattering matrix equations.**

| | | | | |
|---|---|---|---|---|
| **Directional couplers** | Structural Parameters ($i$ = 1 – 7) | Field transmission coefficient $t_i$ | Field cross coupling coefficient $\kappa_i$ | |
| | Scattering matrix equations | $\begin{bmatrix}E_2^+\\E_5^-\end{bmatrix}=\begin{bmatrix}t_1 & j\kappa_1\\j\kappa_1 & t_1\end{bmatrix}\begin{bmatrix}E_1^+\\E_6^-\end{bmatrix},\ \begin{bmatrix}E_1^-\\E_6^+\end{bmatrix}=\begin{bmatrix}t_1 & j\kappa_1\\j\kappa_1 & t_1\end{bmatrix}\begin{bmatrix}E_2^-\\E_5^+\end{bmatrix},$ $\begin{bmatrix}E_4^+\\E_9^-\end{bmatrix}=\begin{bmatrix}t_2 & j\kappa_2\\j\kappa_2 & t_2\end{bmatrix}\begin{bmatrix}E_3^+\\E_{10}^-\end{bmatrix},\ \begin{bmatrix}E_3^-\\E_{10}^+\end{bmatrix}=\begin{bmatrix}t_2 & j\kappa_2\\j\kappa_2 & t_2\end{bmatrix}\begin{bmatrix}E_4^-\\E_9^+\end{bmatrix},$ $\begin{bmatrix}E_8^+\\E_{13}^-\end{bmatrix}=\begin{bmatrix}t_3 & j\kappa_3\\j\kappa_3 & t_3\end{bmatrix}\begin{bmatrix}E_7^+\\E_{14}^-\end{bmatrix},\ \begin{bmatrix}E_7^-\\E_{14}^+\end{bmatrix}=\begin{bmatrix}t_3 & j\kappa_3\\j\kappa_3 & t_3\end{bmatrix}\begin{bmatrix}E_8^-\\E_{13}^+\end{bmatrix},$ $\begin{bmatrix}E_{12}^+\\E_{17}^-\end{bmatrix}=\begin{bmatrix}t_4 & j\kappa_4\\j\kappa_4 & t_4\end{bmatrix}\begin{bmatrix}E_{11}^+\\E_{18}^-\end{bmatrix},\ \begin{bmatrix}E_{18}^-\\E_{11}^+\end{bmatrix}=\begin{bmatrix}t_4 & j\kappa_4\\j\kappa_4 & t_4\end{bmatrix}\begin{bmatrix}E_{12}^-\\E_{17}^+\end{bmatrix},$ $\begin{bmatrix}E_{16}^+\\E_{21}^-\end{bmatrix}=\begin{bmatrix}t_5 & j\kappa_5\\j\kappa_5 & t_5\end{bmatrix}\begin{bmatrix}E_{15}^+\\E_{22}^-\end{bmatrix},\ \begin{bmatrix}E_{15}^-\\E_{22}^+\end{bmatrix}=\begin{bmatrix}t_5 & j\kappa_5\\j\kappa_5 & t_5\end{bmatrix}\begin{bmatrix}E_{16}^-\\E_{21}^+\end{bmatrix},$ $\begin{bmatrix}E_{20}^+\\E_{25}^-\end{bmatrix}=\begin{bmatrix}t_6 & j\kappa_6\\j\kappa_6 & t_6\end{bmatrix}\begin{bmatrix}E_{19}^+\\E_{26}^-\end{bmatrix},\ \begin{bmatrix}E_{19}^-\\E_{26}^+\end{bmatrix}=\begin{bmatrix}t_6 & j\kappa_6\\j\kappa_6 & t_6\end{bmatrix}\begin{bmatrix}E_{20}^-\\E_{25}^+\end{bmatrix},$ $\begin{bmatrix}E_{24}^+\\E_{27}^-\end{bmatrix}=\begin{bmatrix}t_7 & j\kappa_7\\j\kappa_7 & t_7\end{bmatrix}\begin{bmatrix}E_{23}^+\\E_{28}^-\end{bmatrix},\ \begin{bmatrix}E_{23}^-\\E_{28}^+\end{bmatrix}=\begin{bmatrix}t_7 & j\kappa_7\\j\kappa_7 & t_7\end{bmatrix}\begin{bmatrix}E_{24}^-\\E_{27}^+\end{bmatrix}.$ | | |
| **Connecting waveguides** | Structural Parameters ($i$ = 1 – 3) | Length $L_i$ | Transmission factor $a_i$ | Phase shift $\varphi_i$ |
| | Scattering matrix equations | $\begin{bmatrix}E_3^+\\E_2^-\end{bmatrix}=T_1\begin{bmatrix}E_2^+\\E_3^-\end{bmatrix},\ \begin{bmatrix}E_5^+\\E_4^-\end{bmatrix}=T_2\begin{bmatrix}E_4^+\\E_5^-\end{bmatrix},\ \begin{bmatrix}E_7^+\\E_6^-\end{bmatrix}=T_3\begin{bmatrix}E_6^+\\E_7^-\end{bmatrix},\ \begin{bmatrix}E_9^+\\E_8^-\end{bmatrix}=T_2\begin{bmatrix}E_8^+\\E_9^-\end{bmatrix},$ $\begin{bmatrix}E_{11}^+\\E_{10}^-\end{bmatrix}=T_3\begin{bmatrix}E_{10}^+\\E_{11}^-\end{bmatrix},\ \begin{bmatrix}E_{13}^+\\E_{12}^-\end{bmatrix}=T_2\begin{bmatrix}E_{12}^+\\E_{13}^-\end{bmatrix},\ \begin{bmatrix}E_{15}^+\\E_{14}^-\end{bmatrix}=T_3\begin{bmatrix}E_{14}^+\\E_{15}^-\end{bmatrix},$ $\begin{bmatrix}E_{17}^+\\E_{16}^-\end{bmatrix}=T_2\begin{bmatrix}E_{16}^+\\E_{17}^-\end{bmatrix},\ \begin{bmatrix}E_{19}^+\\E_{18}^-\end{bmatrix}=T_3\begin{bmatrix}E_{18}^+\\E_{19}^-\end{bmatrix},\ \begin{bmatrix}E_{21}^+\\E_{20}^-\end{bmatrix}=T_2\begin{bmatrix}E_{20}^+\\E_{21}^-\end{bmatrix},$ $\begin{bmatrix}E_{23}^+\\E_{22}^-\end{bmatrix}=T_3\begin{bmatrix}E_{22}^+\\E_{23}^-\end{bmatrix},\ \begin{bmatrix}E_{25}^+\\E_{24}^-\end{bmatrix}=T_2\begin{bmatrix}E_{24}^+\\E_{25}^-\end{bmatrix},\ \begin{bmatrix}E_{27}^+\\E_{26}^-\end{bmatrix}=T_1\begin{bmatrix}E_{26}^+\\E_{27}^-\end{bmatrix}.$ | | |
| **Input** | | $E_1^+ = 1$, $E_{28}^- = 0$. | | |

## 5. Deviations induced by approximations

As evidenced by the results in previous sections, our method proves effective for modeling integrated photonic resonators and shows advantages for modeling those with complex configurations. Despite this, the method still has limitations that could cause deviations between the simulation results and measurements for practical devices. In this section, we discuss the limitations of our method to model integrated photonic resonators and methods for improving modeling accuracy.

In our modeling, we assume the self-coupling coefficient $t$ and cross-coupling coefficient $\kappa$ of the directional couplers to be wavelength-independent constant, while this holds true only within a certain wavelength range. According to Ref. [34 95 32], the field cross-coupling coefficient $\kappa$ of a directional coupler with a coupling length of $L_c$ can be given by

$$\kappa = \sin\left(\frac{\pi}{2}\cdot\frac{L_c}{L_x}\right), \quad (7)$$

where $L_x$ is the cross-over length, defined as the minimum distance at which optical power completely transfers from one waveguide to the other. The $L_x$ can be further expressed as [95]

$$L_x = \frac{\lambda}{2(n_{eff,\ even}\text{-}n_{eff,\ odd})}. \quad (8)$$

where $\lambda$ is the light wavelength, $n_{eff,\ even}$ and $n_{eff,\ odd}$ are the effective indices of the two fundamental eigenmodes of the coupled waveguides, respectively. From **Eqs. (7)** and **(8)**, it can be seen that the coupling strength of the directional coupler is actually wavelength dependent due to the existence of dispersion (which leads to wavelength-dependent values of $n_{eff,\ even}$ and $n_{eff,\ odd}$). Therefore, the variation in the coupling strength of the directional couplers with wavelength can no longer be ignored when modeling the device's response spectra over a broad bandwidth (typically > 30 nm).

Another potential limitation arises from the fact that we use a constant waveguide group index $n_g$ instead of wavelength-dependent waveguide effective index $n_{eff}$ when calculating the phase shift along waveguides. In fact, the relation between $n_g$ and $n_{eff}$ can be expressed as [94]:

$$n_g = n_{eff}\,(\lambda)\text{-}\ \lambda\frac{dn_{eff}\,(\lambda)}{d\lambda}, \quad (9)$$

where $n_{eff}\,(\lambda)$ is the effective index as a function of light wavelength $\lambda$. The group index $n_g$ is widely used for calculating the free spectral ranges (FSRs) of integrated photonic resonators. For example, the FSR for a MRR can be approximately given by [94]

$$\lambda_{FSR} \approx \frac{\lambda_0^2}{n_g L}, \quad (10)$$

where $\lambda_0$ is the resonance wavelength and $L$ is the ring circumference.

Similar to the cross-coupling coefficient $\kappa$ of a directional coupler, the group index $n_g$ can be regarded as a constant coefficient only within a specific wavelength range. In contrast to introducing $n_{eff}$ at different wavelengths, using a constant $n_g$ can greatly simplify the plotting of the response spectra based on the spectral transfer function. However, this approximation only holds for modeling spectral response over a relatively narrow wavelength range (*e.g.*, < 30 nm). When modeling the spectral response over a broad bandwidth or for nonlinear optical devices that are sensitive to dispersion-induced phase mismatch [8, 96-128], the wavelength-dependent $n_{eff}$ should be used to achieve more accurate simulation results.

## 6. Conclusion

In summary, a universal approach for modeling integrated photonic resonators with complex structures is proposed based on the SMM. By dividing the device configuration into basic elements such as directional couplers and connecting waveguides, our approach shows effectiveness in modeling integrated photonic resonators with both unidirectional and bidirectional light propagation, with the simulated spectral response agreeing well with experimental results. By dividing the device configuration into independent submodules MRRs and SIs, the modeling in our approach can also be simplified. Finally, we discuss the limitations arising from approximations in our modeling and the corresponding strategies for improving modeling accuracy. Our approach offers an efficient way for designing and optimizing complex integrated photonic devices, which is applicable across a wide range of integrated platforms. [129-147]

**Conflicts of Interest:** The authors declare no conflicts of interest.